# Digital libraries: textual analysis for a systematic review and meta-analysis

Mathieu Andro (1), Marc Maisonneuve (2)
(1) bibliotheque-numerique.fr - mathieuandro@yahoo.fr - Paris, France
(2) toscaconsultants.fr - marc.maisonneuve@toscaconsultants.fr - Paris, France

# Abstract

**Purpose**: We seek to explore the realm of literature about digital libraries. We specifically seek to ascertain how interest in this subject has evolved, its impact, the most productive journals and countries, the number of occurrences of digital libraries, the relationships and dynamics of the main concepts mentioned, and the dynamics of metadata formats.
**Methods**: We extracted corpora from the Google Scholar and Microsoft Academic Search bibliographic databases. We analyzed the named entities and concepts contained within these corpora with the help of text mining technologies, CorTexT in particular.
**Results**: While the number of publications on the subject of digital libraries is increasing, their average number of citations is decreasing. China, the United States and India are the most productive countries on the subject. Literature about conservation and national libraries has gradually been replaced by literature about open access, university libraries and the relationship with users. Internet Archive is the most cited digital library in literature and continues to grow. Dublin Core is the most talked about metadata format, however the subject of metadata formats is declining in the corpus today.
**Conclusion**: Digital libraries now seem to be reaching the age of maturity.

**Keywords**: Digital libraries; Digitization; Text mining; CorTexT; Systematic review; Meta-analysis

# Declaration

This research was not funded.
We work on consulting missions and also for the French government.
All the data we have used can be found freely and free of charge online on Google Scholar and Microsoft Academic Search. Text analyses can also be reproduced with the free CorTexT text mining software.

# 1. Introduction. Literature About Digital Libraries

## 1.1. The dynamics of literature about digital libraries

Most scientific and professional digital literature is indexed in Google Scholar. We therefore chose to use this resource for our initial analyses. The first question we asked ourselves was whether the subject of digital libraries was generating a lot of ink, and particularly what the dynamic was over time. To do this, we simply searched Google Scholar with the query "digital library" OR "digital libraries" for each year since 2000, counting the number of publications for the year. We then built a histogram using a spreadsheet in order to identify the major trends.

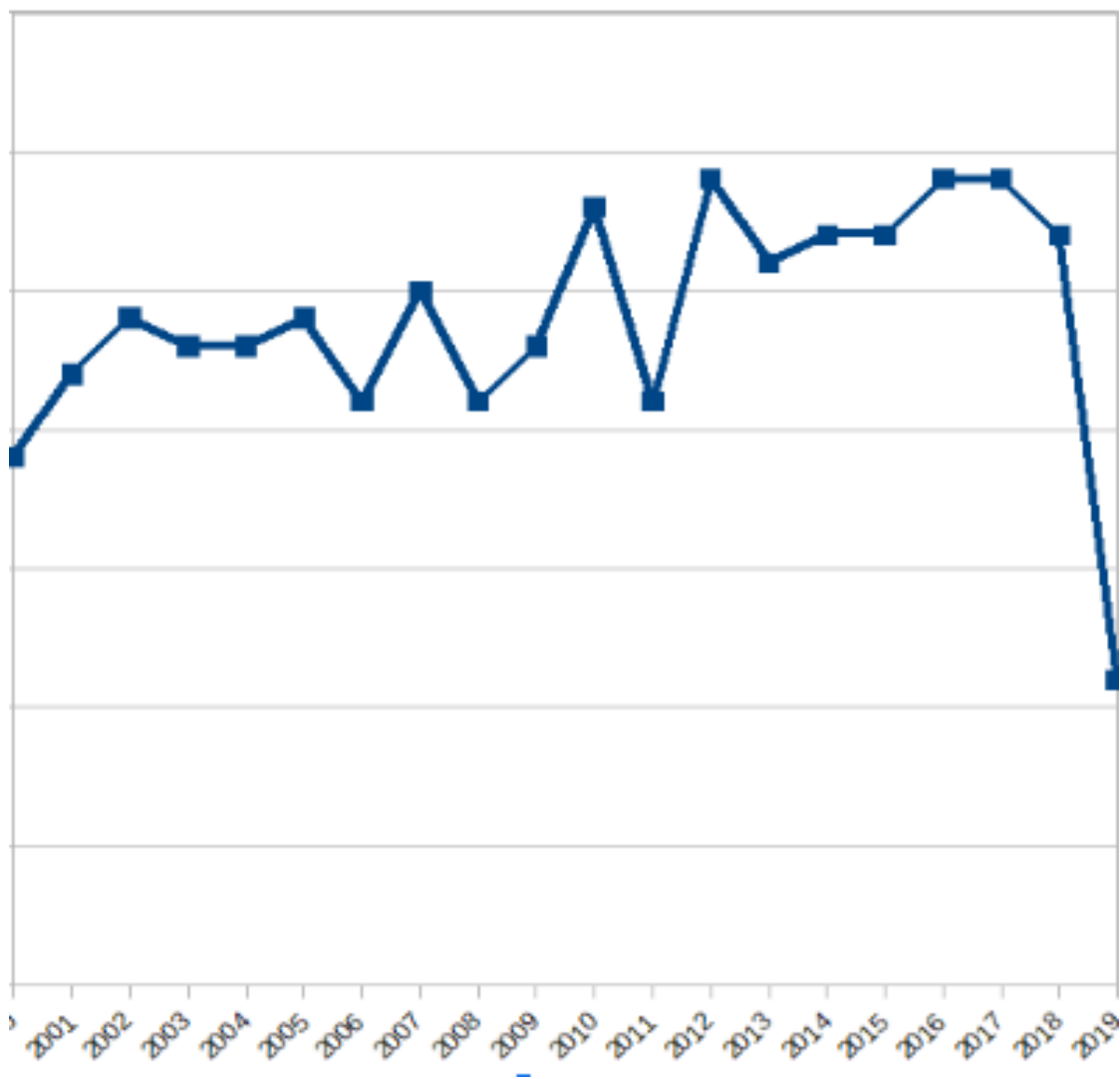

**Graphic 1. Evolution of the number of publications referenced by Google Scholar corresponding to the query "digital library" OR "digital libraries"**

In total, Google Scholar identifies around 19,600 books or articles published over the period, i.e. an average of slightly less than 1,000 publications per year. Surprisingly, the number of publications in 2019 is at odds with previous years.. It is quite likely that in April 2020, when these analyses were produced, not all publications for 2019 had yet been indexed in Google Scholar

We observe overall growth in worldwide literature produced on the subject of digital libraries in the early 2000s, followed by a decline from 2017 onwards. With digital libraries having moved well beyond the research and development stage, they now seem to be of less interest to scientific research.. In a way, digital libraries seem to have reached the stage of maturity.

## 1.2. Corpus extraction methods and tools

The Google Scholar corpus cannot be exported for security reasons and therefore does not allow filtering by keywords as we would have liked, in order to analyze the literature in more depth. We therefore used Harzing's Publish or Perish software to extract a corpus of 19,737 bibliographic references on the "Microsoft Academic" bibliographic database from our query on the keywords "digital libraries" OR "digital library".

## 1.3. Impact in number of citations

We also sought to ascertain in our analyses the impact of this literature through citations in bibliographies within literature. We obtained an average of 5.21 citations per publication. We then sought to determine the evolution of the average of this number of citations according to the year of publication:

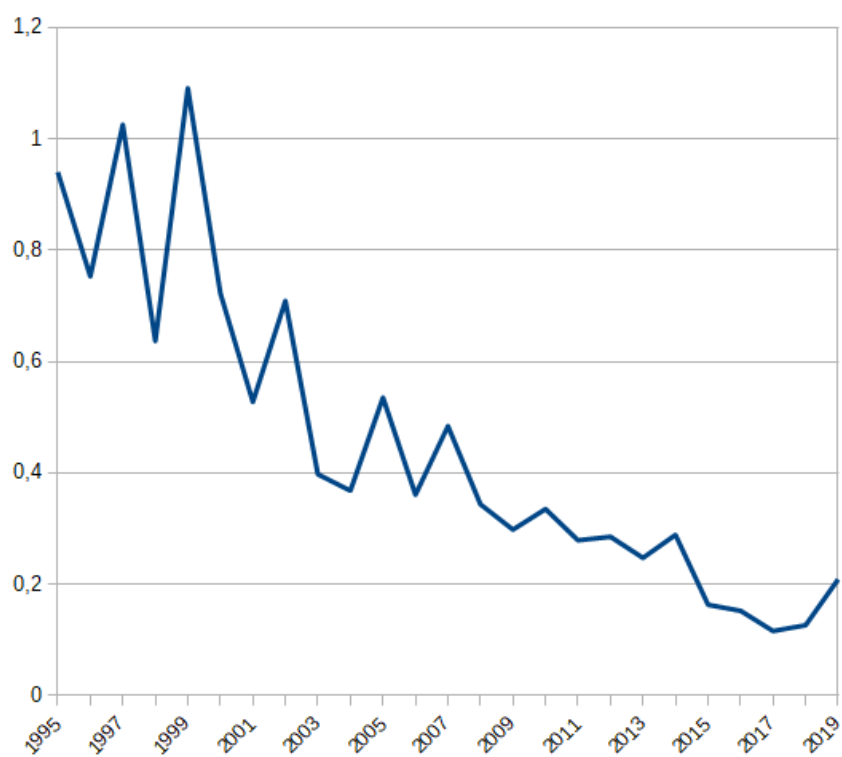

**Graph 2. Evolution of the average number of citations per year of publications in the Microsoft Academic corpus corresponding to the query "digital library" OR "digital libraries"**

This number has steadily declined. This means that publications on the subject of digital libraries are having a diminishing impact on literature. Digital libraries are therefore no longer a growing topic of research.

# 2. Results

## 2.1. Who publishes on digital libraries?

### 2.1.1. Which journals?

Tree structure of the main journals most represented in the corpus, created using Tableau Public software

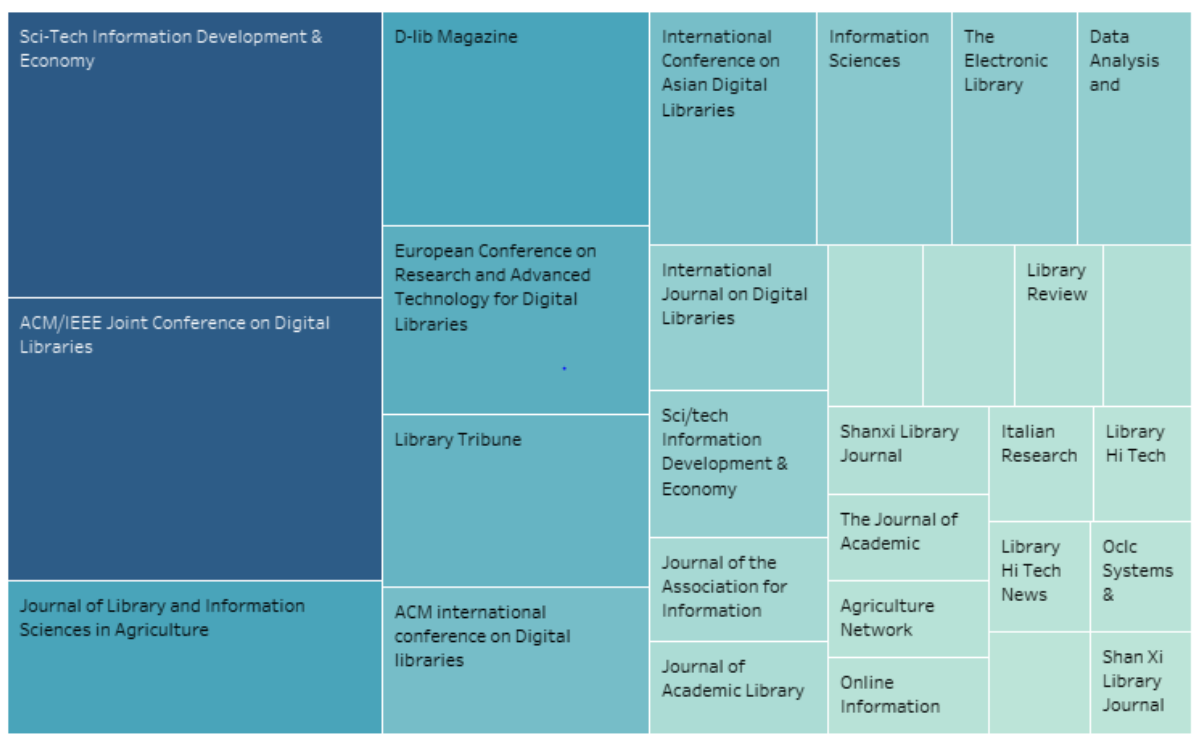

**Graph 3. Main journals most represented in the corpus**

**Source**: Microsoft Academic
**Query**: "digital library" OR "digital libraries"
**Corpus**: 19,703 references
**Diagram**: the surface of each rectangle is proportional to the number of bibliographic references for each journal.

There are a large number of conference publications in the corpus. Next in number are classic journals in information sciences, with the most important impact factors which are, unsurprisingly, the most represented in the corpus.
 Among the diverse sources represented in the Microsoft corpus, we observe a strong presence of articles from the journals Sci-Tech Information Development & Economy, ACM /

IEEE Joint Conference on Digital Libraries, Journal of Library and Information Sciences in Agriculture and D- lib Magazine.

### 2.1.2. Which countries?

We decided to consider only the most frequent geographic named entities. As these are entities of various specific levels (cities, regions, countries, continents), we also decided to consider country level only. In order to visualize the impact of countries on literature, we used a recognition algorithm of geographic named entities from CorTexT, on the title and summary text fields, and we visualized the number of occurrences per country via http://www.mapinseconds.com:

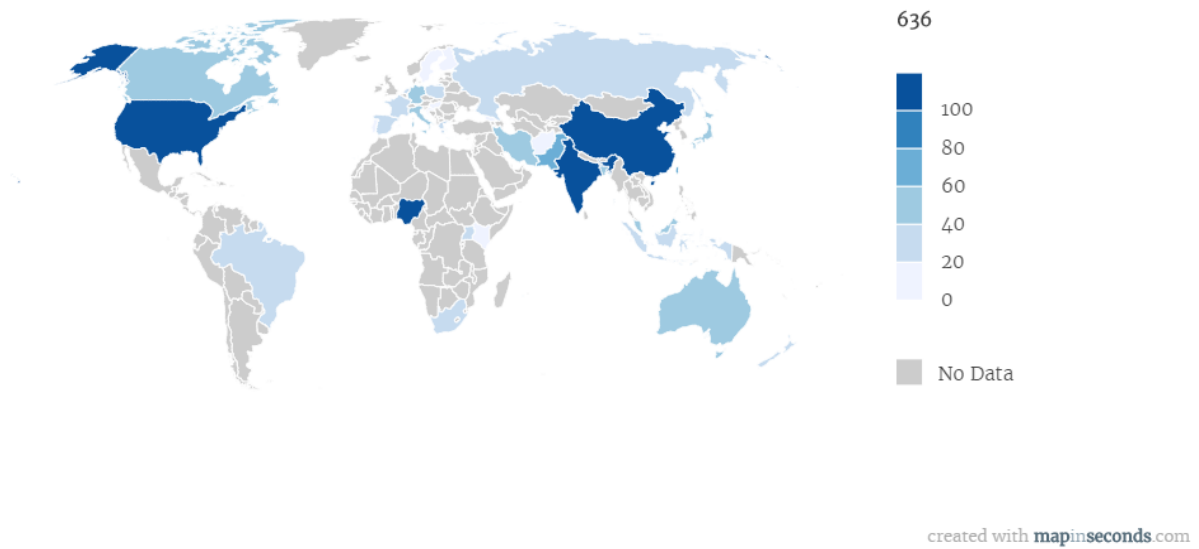

**Graphic 4. Main countries most represented in the corpus**

We observe that China clearly dominates with 636 occurrences in the corpus, followed by the United States with 314 occurrences, then 278 for India, 152 for the United Kingdom and 104 for Nigeria.

## 2.2. What is this literature about?

### 2.2.1. Concepts map

We used https://www.cortext.net to obtain text analysis of the corpus.

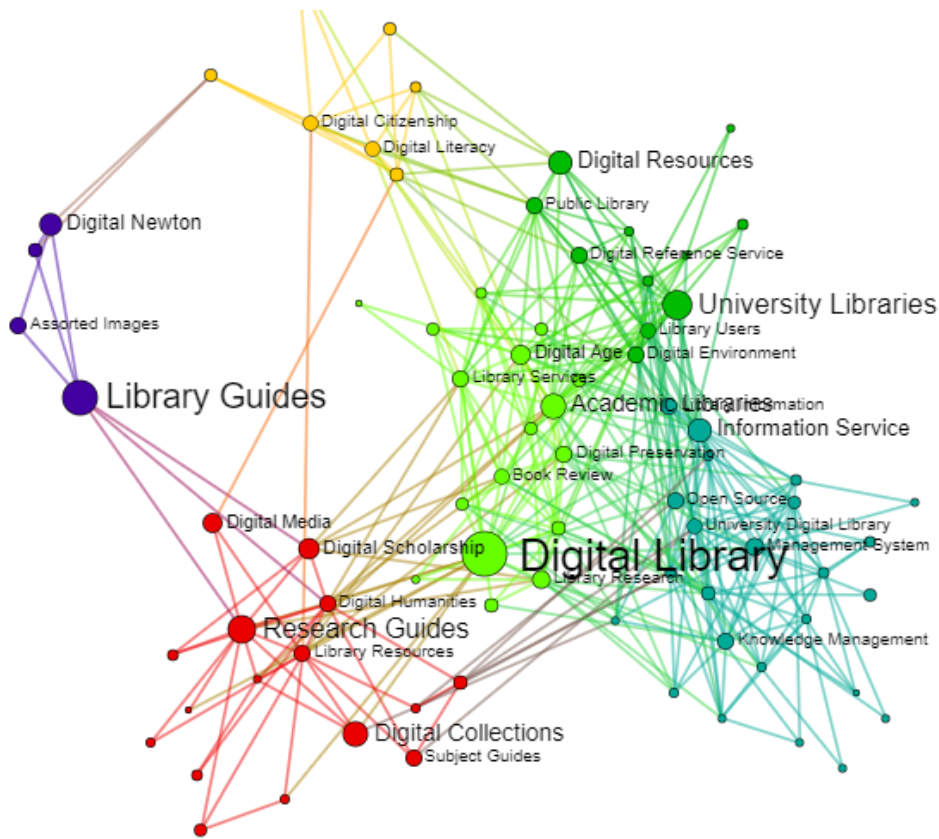

**Graph 5. Map of concepts represented in the corpus**

**Source**: Microsoft Academic
**Query**: "digital library" OR "digital libraries" + date> 2000
**Corpus**: 19,703 references
**Diagram**: grouping by co-occurrences of the most frequent keywords (script network mapping)
**Community detection algorithm**: Louvain
**Fields**: terms
**Nodes**: 150

Data driven analysis of the themes, i.e. without initial assumptions or prejudices, allows us to identify management and conservation law issues that are different from service and usage analysis issues, which are themselves distinct from guides for university audiences in particular.

## 2.2.2. History of the literature about digital libraries

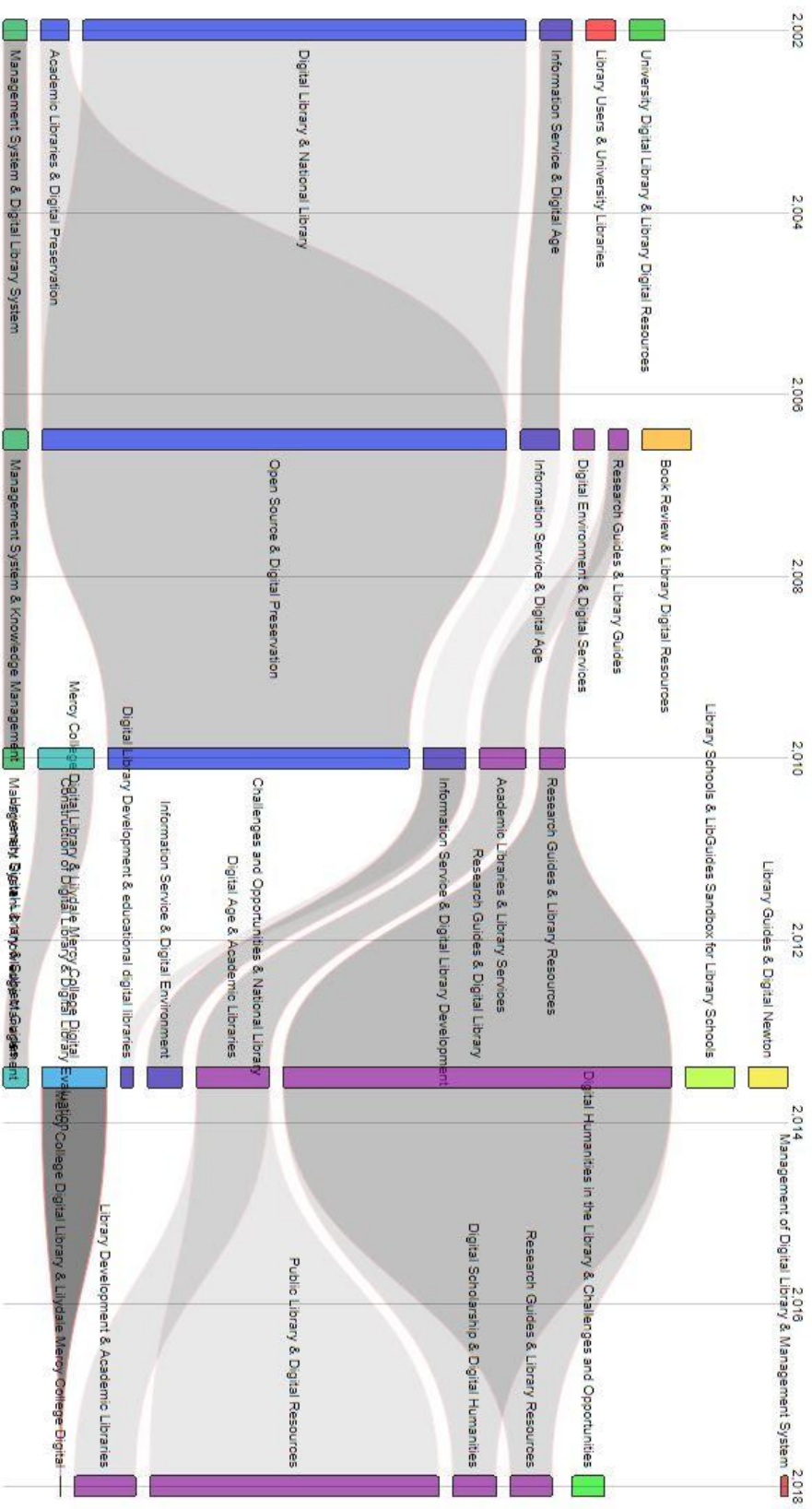
8

**Graph 6. Dynamics of the importance of concepts represented in the corpus**

**Source**: Microsoft Academic
**Query**: "Digital library" OR "digital libraries" + date> 2000
**Corpus**: 19,703 references
**Diagram**: grouping by co-occurrences of the most frequent keywords (script network mapping)
**Community detection algorithm**: Leuven
**Fields**: terms
**Nodes**[1]: 150

Until 2008, we see a strong predominance of subjects related to conservation policies and national libraries over other themes such as university libraries, knowledge management or reflections on the information society. Somewhat late, from 2005 to 2010, we also see significant growth in literature produced on the subject of digital university libraries, albeit still from a preservation point of view..
Around 2008, we see current status reports and guides emerge, the production of which increases until 2014, when they even represent the bulk of the corpus before waning to the present day. Between 2010 and 2015, we also see the emergence of literature on the evaluation of digital libraries. And lastly, we are also able to observe the emergence of the notion of digital humanities since 2016 within the corpus of digital libraries. Reflection on the information society decreased until 2008 with the emergence of the digital library as a service and even of its relationship with citizens from 2012 onwards. The user's perspective now seems to be the main subject still in development.

## 2.2.3. Large digital libraries in literature

After these initial analyses, we also thought it would be interesting to compare the presence of large digital libraries in literature over time. We used the same method to do this, querying the names of these libraries in Google Scholar for each year since 2000 and analyzing the results using the following diagram produced with spreadsheet software.

---
[1] Number of nodes in the network

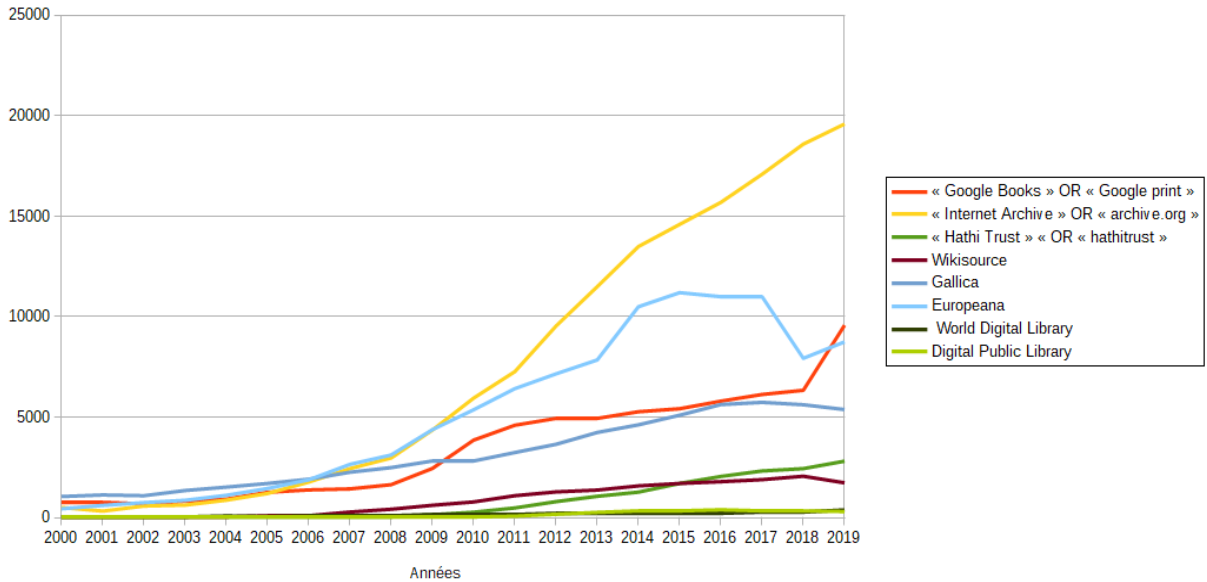

**Graph 7. The presence of the main digital libraries
in literature over the past 20 years according to Google Scholar**

If we compare the number of occurrences of the main digital libraries in literature via Google Scholar, we see that Archive.org (Internet Archive) is the digital library most mentioned in literature followed by Europeana. It is mentioned more than Google Books, which paradoxically offers more content, and has done so since 2005. It is also cited more often in literature than Gallica since 2007. The number of occurrences of Gallica decreased slightly from 2018 and those of Wikisource from 2019.

For the remainder of our analyses, we therefore chose to focus on archive.org (Internet Archive), Google Books, Hathi Trust Digital Library, Gallica and Wikisource, leaving aside Europeana which is a metadata aggregator and not a digital library. Europeana therefore does not offer a direct distribution solution for content digitized by a library. We have also left aside the World Public Library and the Digital Public Library which occupy a more modest ranking in the previous classification.

2.2.4. What about metadata?

**Distribution of occurrences of mentions of metadata formats throughout Google Scholar over the past 20 years (raw data)**

| Format | 2000 | 2001 | 2002 | 2003 | 2004 | 2005 | 2006 | 2007 | 2008 | 2009 | 2010 | 2011 | 2012 | 2013 | 2014 | 2015 | 2016 | 2017 | 2018 | 2019 |
|---|---|---|---|---|---|---|---|---|---|---|---|---|---|---|---|---|---|---|---|---|
| Dublin Core[2] | 6250 | 7280 | 7860 | 8350 | 9480 | 10200 | 11000 | 12200 | 13700 | 14800 | 16200 | 18600 | 19800 | 22800 | 23400 | 25800 | 22800 | 21800 | 23500 | 21500 |
| EAD[3] | 120 | 155 | 201 | 253 | 260 | 271 | 261 | 260 | 263 | **263** | 232 | 311 | 313 | 337 | 267 | 255 | 244 | 294 | 222 | 159 |
| MARC[4] | 105 | 110 | 136 | 158 | 184 | 181 | 181 | 213 | 230 | 269 | 329 | 386 | 376 | 384 | 394 | 343 | 324 | 387 | 241 | 282 |
| METS[5] | 1 | 6 | 61 | 86 | 120 | 163 | 183 | 184 | 186 | 200 | 184 | 194 | 191 | 194 | 141 | 143 | 168 | 150 | 145 | 78 |
| MODS[6] | 0 | 0 | 20 | 37 | 64 | 103 | 131 | 122 | 116 | 135 | 150 | 162 | 160 | 198 | 160 | 144 | 146 | 173 | 130 | 96 |

**Table 1. Distribution of occurrences of mentions of semantic web formats in Google Scholar over the past 20 years (raw data)**

| Format | 2000 | 2001 | 2002 | 2003 | 2004 | 2005 | 2006 | 2007 | 2008 | 2009 | 2010 | 2011 | 2012 | 2013 | 2014 | 2015 | 2016 | 2017 | 2018 | 2019 |
|---|---|---|---|---|---|---|---|---|---|---|---|---|---|---|---|---|---|---|---|---|
| FRBR[7] | 1 | 1 | 115 | 1 | 1 | 248 | 2 | 319 | 1 | 2 | 398 | 511 | 520 | 2 | 510 | 467 | 1 | 479 | 365 | 279 |
| IFLA LRM[8] | 0 | 0 | 0 | 0 | 0 | 0 | 0 | 0 | 0 | 0 | 0 | 0 | 0 | 0 | 0 | 0 | 1 | 63 | 126 | 89 |
| RDA[9] | 1 | 1 | 2 | 4 | 12 | 43 | 83 | 152 | 204 | 268 | 300 | 422 | 432 | 526 | 492 | 515 | 474 | 460 | 409 | 325 |
| RDF[10] | 7 02 | 975 | 1450 | 2000 | 2530 | 2710 | 3050 | 3070 | 3260 | 3560 | 3820 | 3990 | 4150 | 4310 | 4260 | 4310 | 4240 | 4130 | 3940 | 3000 |
| SKOS[11] | 0 | 2 | 4 | 4 | 5 | 27 | 37 | 56 | 157 | 291 | 392 | 385 | 449 | 504 | 495 | 531 | 462 | 478 | 410 | 410 |
| Sparql[12] | 124 | 154 | 184 | 164 | 233 | 495 | 1120 | 1880 | 2760 | 3340 | 4010 | 4570 | 5310 | 5880 | 6250 | 6250 | 6210 | 5710 | 5900 | 5350 |
| TEI[13] | 319 | 197 | 358 | 335 | 386 | 395 | 437 | 402 | 449 | 511 | 494 | 501 | 589 | 615 | 656 | 613 | 788 | 673 | 654 | 574 |

**Table 2. Graphical distribution of the occurrences of mentions of metadata formats in Google Scholar over the past 20 years**

---

[2] Query: "Dublin Core"
[3] Query: "Encoded Archival Description"
[4] Query: "machine readable cataloging"
[5] Query: "Metadata Encoding and Transmission Standard"
[6] Query: "Metadata Object Description Schema"
[7] Query: "Functional Requirements for Bibliographic Records"
[8] Query: "IFLA LRM"
[9] Query: "Resource Description and Access"
[10] Query: "Resource Description Framework"
[11] Query: "Simple Knowledge Organization System"
[12] Query: Sparql
[13] Query: "text encoding initiative"

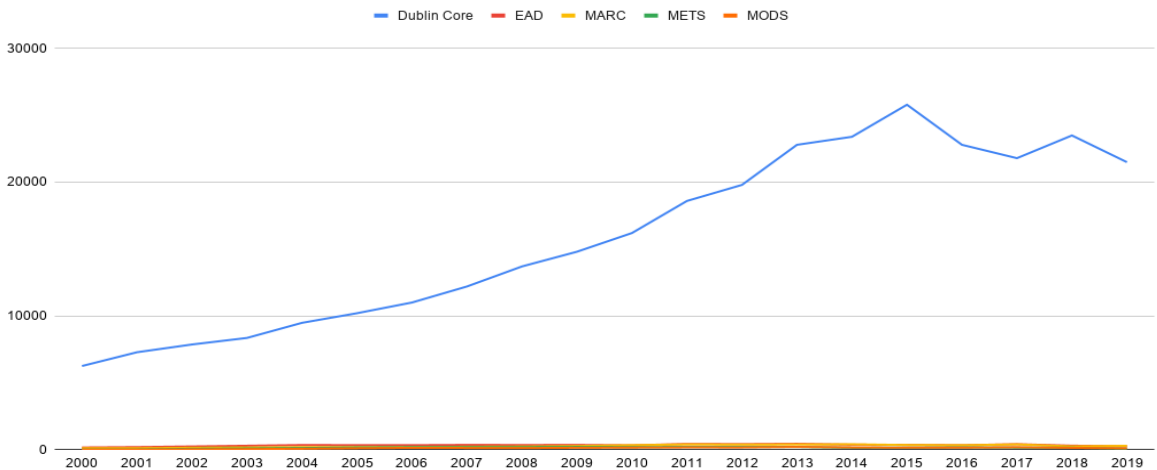

**Graph 8. Dynamics of the importance of metadata represented in the corpus**

**Source and corpus**: Google Scholar
**Queries**: "Dublin Core"; "Encoded Archival Description"; "machine readable cataloging"; "Metadata Encoding and Transmission Standard"; "Metadata Object Description Schema"
**Diagram**: number of publications per year since 2000
**Comment**: This diagram shows that Dublin Core is very dominant compared to other metadata formats. This format was initially defined in 1995, in 2003 for the ISO 15836 standard, then 2009 for the new version of this standard and finally 2017 and 2019: https://www.iso.org/fr/search.html?q= 15836). There are other dates for 2002 and 2003 but less important:
- Expressing Simple Dublin Core in RDF / XML: http://dublincore.org/documents/2002/07/31/dcmes-xml/
- Guidelines for implementing Dublin Core in XML: http://dublincore.org/documents/2003/04/02/dc-xml-guidelines/
- Expressing Qualified Dublin Core in RDF / XML: http://dublincore.org/documents/2002/05/15/dcq -rdf-xml /]

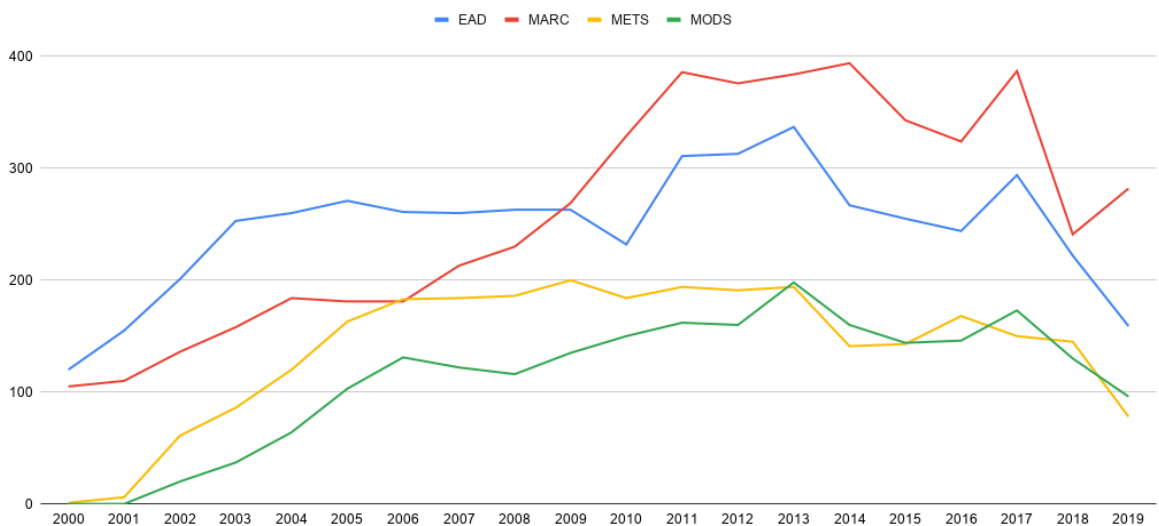

**Graph 9. Graphic distribution of the occurrences of mentions of metadata formats (except Dublin Core) in Google Scholar over the past 20 years**

**Source and corpus**: All Google Scholar
**Queries**: "Encoded Archival Description"; "MAchine Readable Cataloging"; "Metadata Encoding and Transmission Standard"; "Metadata Object Description Schema"
**Diagram**: Number of publications per year since 2000
**Comments**:
- The MARC format was defined in 1965
- The EAD format was defined in 1998 for V1, in 2002 for V2 and 2014 for V3
- The METS format was defined in April 2001
- The MODS format was defined in 2002

We observe an increase in mentions of these formats until 2011 followed by a decline from 2018. This dynamic is quite similar to the time curve for the entire corpus of digital libraries. The dynamic curves of all these formats are quite parallel. For example, peaks were observed for each format in 2013 and 2017. It is likely that these formats are often mentioned in joint publications. We also observe a strong resilience of MARC formats and a lesser interest in MODS in literature since this format is located in an intermediate position between Dublin Core and MARC.

Regarding metadata, Dublin Core obviously occupies a hegemonic position, well ahead of MARC, followed by EAD, then MOD and then METS. We also observe a decline in the number of occurrences in literature for all of these formats: from 2009 for METS, since 2013 for EAD and METS and since 2014 for MARC.

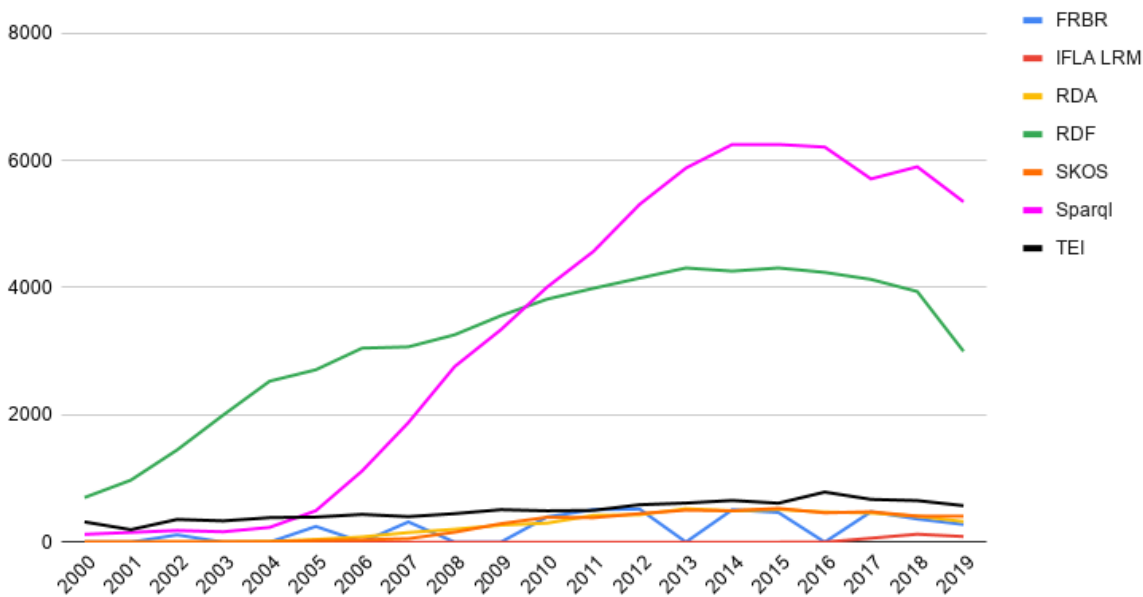

**Graph 10. Graphic distribution of the occurrences of mentions of semantic web formats in Google Scholar over the past 20 years**

**Source and corpus**: All Google Scholar
**Queries**: "" Functional Requirements for Bibliographic Records ";" IFLA LRM ";" Resource Description and Access ";" Resource Description Framework ";" Simple Knowledge Organization System "; Sparql;" Text Encoding Initiative "
**Diagram**: Number of publications per year since 2000
**Comments**: Semantic web formats seem to generate more interest in literature than metadata. However, we have also observed the same decline in recent years as for metadata.

Regarding semantic web, we see a greater number of occurrences in Google Scholar for Sparql followed by RDF, TEI, SKOS, RDA, FRBR then IFLA LRM. We notice with astonishment a very irregular and jagged presence of quotations from the FRBR format in literature followed by a decline since 2012. This decline in the number of occurrences can also be observed for RDA since 2013, for RDF, SKOS and Sparql since 2015 and for TEI since 2016.

This decline in metadata and semantic web formats is probably related to the decline in volume of literature produced on the subject of digital libraries since 2017, as previously mentioned.

# Conclusion

With just under 20,000 publications on the subject since 2000, digital libraries now seem to be reaching the age of maturity. While the number of publications is now on a downward trend, as is the number of citations per year, certain projects - archive.org, Google Books or even Europeana - seem to be enjoying unabated interest. China, well ahead of the United

States and India, is now the country most mobilized on the issue of digital libraries. The issue of uses, evaluation of projects and even digital humanities seem to have taken precedence over technical aspects, conservation policies, current status reports and the semantic web.